\documentclass[a4paper,11pt]{article}
\usepackage{pos}
\usepackage{graphicx}

\title{EUSO-SPB2 Fluorescence Telescope in-flight performance and preliminary results }
 \ShortTitle{EUSO-SPB2 FT}
 \author*[a]{G. Filippatos}
 \affiliation[a]{Colorado School of Mines, Golden, USA}
 \emailAdd{gfilippatos@mines.edu}
 \onbehalf{for the JEM-EUSO Collaboration\\[-1mm]{\normalsize \normalfont
(a complete list of authors can be found at the end of the proceedings)}}

\abstract{
	The Extreme Universe Space Observatory on a Super Pressure Balloon II (EUSO-SPB2) launched from Wanaka, New Zealand on May 13$^{\text{th}}$ 2023. 
	Consisting of two optical telescopes, EUSO-SPB2 aimed to search for very high energy neutrinos (E>PeV) via Cherenkov radiation, and ultra high energy cosmic rays (UHECRs, E>EeV) via ultraviolet fluorescence. 
	Building on the EUSO-Balloon (2014) and EUSO-SPB1 (2017) missions, the Fluorescence Telescope (FT) comprises 108 multi-anode photomultiplier tubes at the focus of a one meter entrance diameter Schmidt telescope. 
	The FT pointed down at the atmosphere below the SPB’s altitude of 33 km. 
	The mission duration was planned to reach up to 100 days. 
	Prior to flight, the instrument was extensively tested in the laboratory and in the field. 
	These measurements, combined with simulations led to an expected peak energy sensitivity around 3 EeV. 
	Combined with a three-times-larger field-of-view than previous EUSO balloon missions, this resulted in an expected observation rate of one UHECR shower per ten hours of observation. 
	The FT was expected to perform the first measurement of UHECRs via fluorescence from sub-orbital space, but was unable to, due to a shortened flight. 
	Nonetheless, these observations of EUSO-SPB2 serve as a stepping stone to future satellite-based missions, such as the Probe of Extreme Multi-Messenger Astrophysics (POEMMA), with enormous exposure to the highest energy cosmic rays with all sky coverage.
	In this contribution we will discuss the performance of the FT in flight as well as preliminary results. 
}

\ConferenceLogo{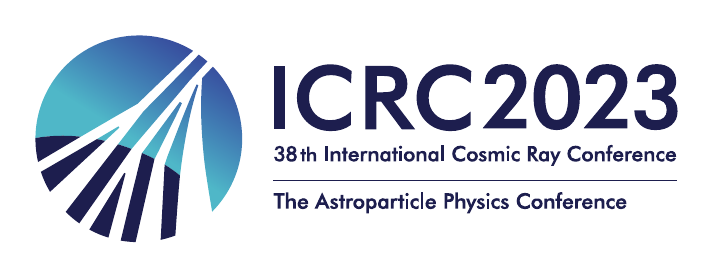}

\FullConference{
38th International Cosmic Ray Conference (ICRC2023)\\
  26 July - 3 August, 2023\\
  Nagoya, Japan
}

\begin{document}
\maketitle

\section{Instrument Overview}

\begin{figure}[h!]\begin{centering}
	\includegraphics[width=0.65\paperwidth]{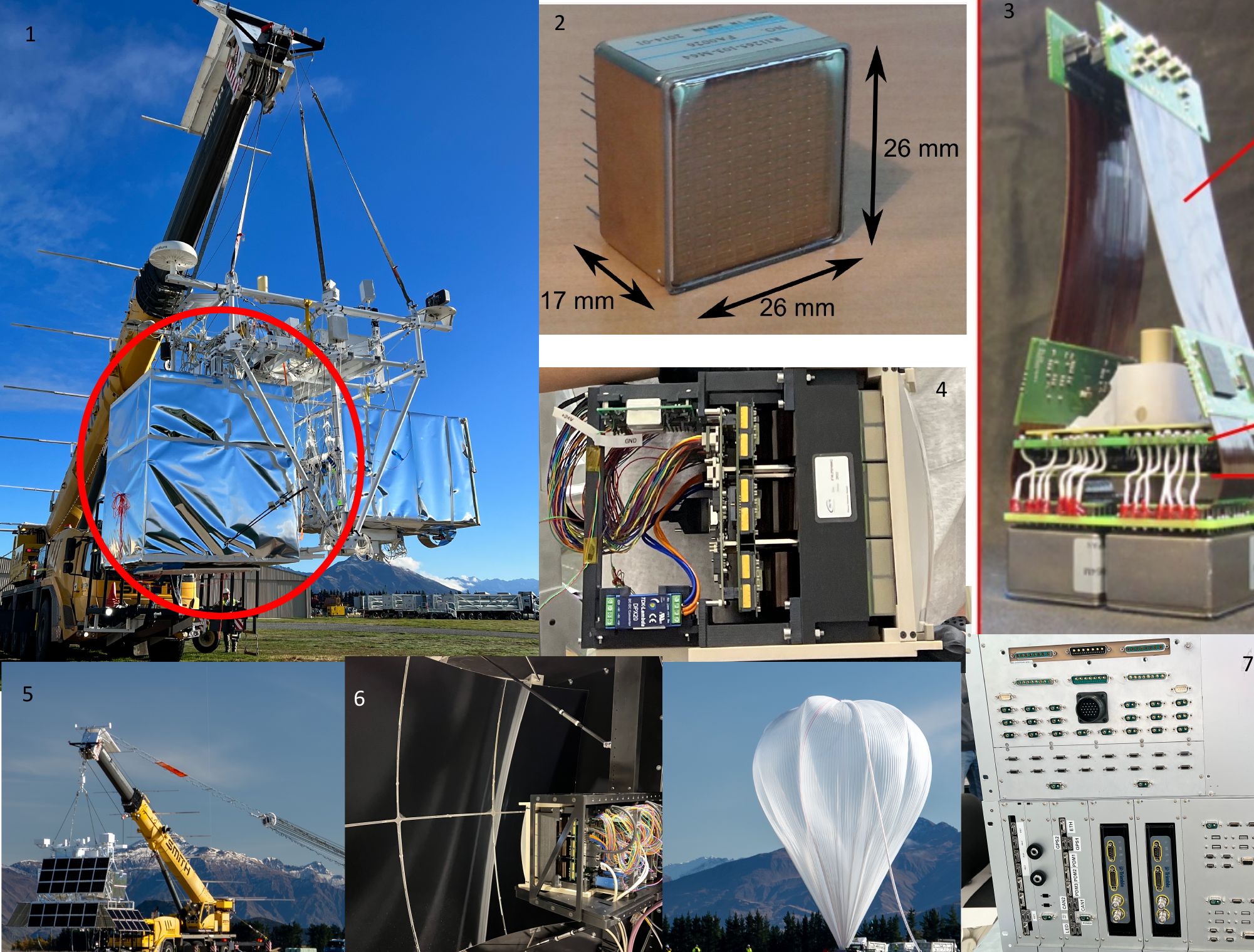}
	\caption{Fully assembled payload, without solar panels, hanging on launch vehicle (1). 
	Individual MAPMT (2).
	Assembled EC, prior to potting, showing HVPS control board and ASIC chips (3).
	Assembled PDM, including BG3 filter and field flattener (4).
	Fully assembled payload on launch vehicle and balloon during inflation (5).
	View of the inside of the telescope taken during integration (6).
	Data processor and power distribution crates before assembly (7). }
	\label{figs:pics}
\end{centering}
\end{figure}

The Fluorescence Telescope (FT) on board EUSO-SPB2 built on the experience of previous EUSO missions, and was a stepping stone towards future space based missions such as the Probe of Extreme Energy Astrophysics (POEMMA) \citep{POEMMA}.
EUSO-Balloon \citep{EUSO-Ballon}, a one night flight from Timmins Canada in 2015, and EUSO-SPB1 \citep{SPB1}, a planned long duration flight from Wanaka New Zealand in 2017, each flew one photo-detection module (PDM) and refractive optics.
The FT was comprised of three PDMs at the focus of a modified Schmidt telescope. 
The various pieces of the FT, along with the fully assembled payload are shown in Figure \ref{figs:pics}.

Each PDM is made up of 36 multi-anode photo-multiplier tubes (MAPMTs, Hamamatsu Photonics R11265-M64) with 64 pixels each. 
Four MAPMTs are grouped together and potted in a gelatinous compound to form an Elementary Cell (EC) with a shared high voltage power supply (HVPS) utilizing a Cockroft-Walton circuit. 
Each MAPMT is read out by a SPACIROC3 ASIC \citep{BLIN2018363} performing both photon counting and charge integration (so-called KI).
The KI allows for measurements of bright signals concentrated in a short pulse outside of the dynamic range of the photon-counting channel. 
The photon counting channel of the 2,304 total pixels are digitized with a 952 kHz frequency and a double pulse resolution of $\sim$6 ns.
While the KI channel of the MAPMTs are digitized at the same cadence, 8 photon counting pixels are grouped together, resulting in 8 KI channels per MAPMT.
Data from the 36 ASICs are sent to three cross boards containing an Artix 7 FPGA which multiplex the data.
The PDM is controlled by a Xilinx Zynq 7000 FPGA with an embedded dual core ARM9 CPU on a custom readout board which handles triggering and data packaging among other tasks \citep{SPB2Trigger}. 

A data processor containing redundant CPUs, redundant differential GPSs, a housekeeping board and a clock board, controls the PDMs. 
The clockboard synchronizes the readout of the three PDMs by receiving the output of the trigger logic on each Zynq board and issuing signals to the three PDMs in parallel. 
Additionally, data from two differential GPS are packaged with trigger and deadtime information for each event that is recorded by the clockboard. 
The CPU handles the commanding of all subsytems as well as the combining of data from the three PDMs and the clockboard. 
In addition to internal commanding, the CPU receives and processes commands from ground. 
Further, the CPU compresses data for download and transmits all monitoring data to ground. 
Monitoring data includes 18 temperature probes, which were placed around the payload, humidity and pressure sensors, gyroscope information, and data from two photo-diodes at the focus of the telescope. 
There are two calibrated health-LEDs (HLEDs) beneath the PDMs which allow the response of the instrument to be measured at four fixed intensities. 

The three PDMs are at the focus of a 1-m diameter entrance pupil Schmidt telescope, behind field flatteners and BG-3 filters. 
Comprised of six segmented spherical mirrors and an aspheric corrector plate, the telescope focuses light into a spot size of $\sim$2 mm with an optical throughput of 67\%, far greater than refractive optics flown on previous EUSO missions. 
The FT is mounted in one half of the EUSO-SPB2 gondola, beneath the Science Information Package (SIP).
Three independent telemetry links are handled by the SIP, with a fourth, a Starlink connection, connected directly to the FT CPU.  
Next to the FT, and  beneath the gondola control computer and science power system, hangs the Cherenkov telescope \citep{ct}, with a tilt mechanism allowing for +5/-13 degrees of pointing with respect to horizontal.
The 5,025 lbs of instrumentation and support equipment, with an additional 600 lbs of ballasts, were suspended from an 18.8 million ft$^3$ (532,000 m$^3$) super pressure balloon.

\section {Pre-Flight Characterizations}

Before the instrument was fully assembled, the PDMs were extensively characterized in the lab using calibrated light sources. 
The telescope was characterized using a 1-m parallel beam system and photodiodes. 

Prior to flight, in August of 2022, the instrument was field tested in Delta, Utah, outside of the Telescope Array Black Rock Mesa Fluorescence Detector. 
The optical signature from extensive air showers (EAS) was mimicked by a pulsed frequency tripled Nd-YAG laser at 355 nm, with a robust automated pointing system. 
Calibrated LEDs placed both in the near and far fields were used to provide an alternative calibration to the piecewise method used in the lab.
Night sky background was observed for tens of hours, including stars, meteors and airglow. In the weeks before flight, the telescope was tested again in Wanaka, New Zealand. 
The same calibrated LEDs used in Delta were utilized and the night sky was observed.

\section{Flight Summary}

After launch at 00:02 UTC on May 13$^{\text{th}}$ 2023, EUSO-SPB2 had a nominal ascent. 
All systems, except for the HVPS, were powered starting before launch and throughout the flight. 
Housekeeping monitoring worked as expected. 
Above 18 km altitude, an oversight of a firmware limitation made our GPS un-usable.
After launch, there were stability issues with the highest bandwidth telemetry connection, which limited our download capacity and increased latency in the communication with the instrument. 
This, however, was fixed in the following day, and did not inhibit our ability to operate the instrument even during this first night, which was devoted to commissioning.

Upon initializing the first data taking at float altitude, it appeared that not all ECs were able to be powered at the nominal HV of 1013 V. 
Re-initializing the instrument several times led to different ECs reaching nominal voltage, with one third to one half of the focal surface being active. 
This was indicative of the good health of the ECs themselves, but also of the necessity to modify the powering scheme in flight conditions. 
A slower increase of the HV during the powering sequence was adopted the second night, which indeed proved fully satisfactory, with all ECs turned on at the intended potential difference, also with nominal efficiency.
Data taking continued under these conditions for the remainder of the dark period on May 13$^{\text{th}}$, with plans to diagnose the HVPS issues the following night, using full telemetry.

\begin{figure}[h!]\begin{centering}
	\includegraphics[width=0.65\paperwidth]{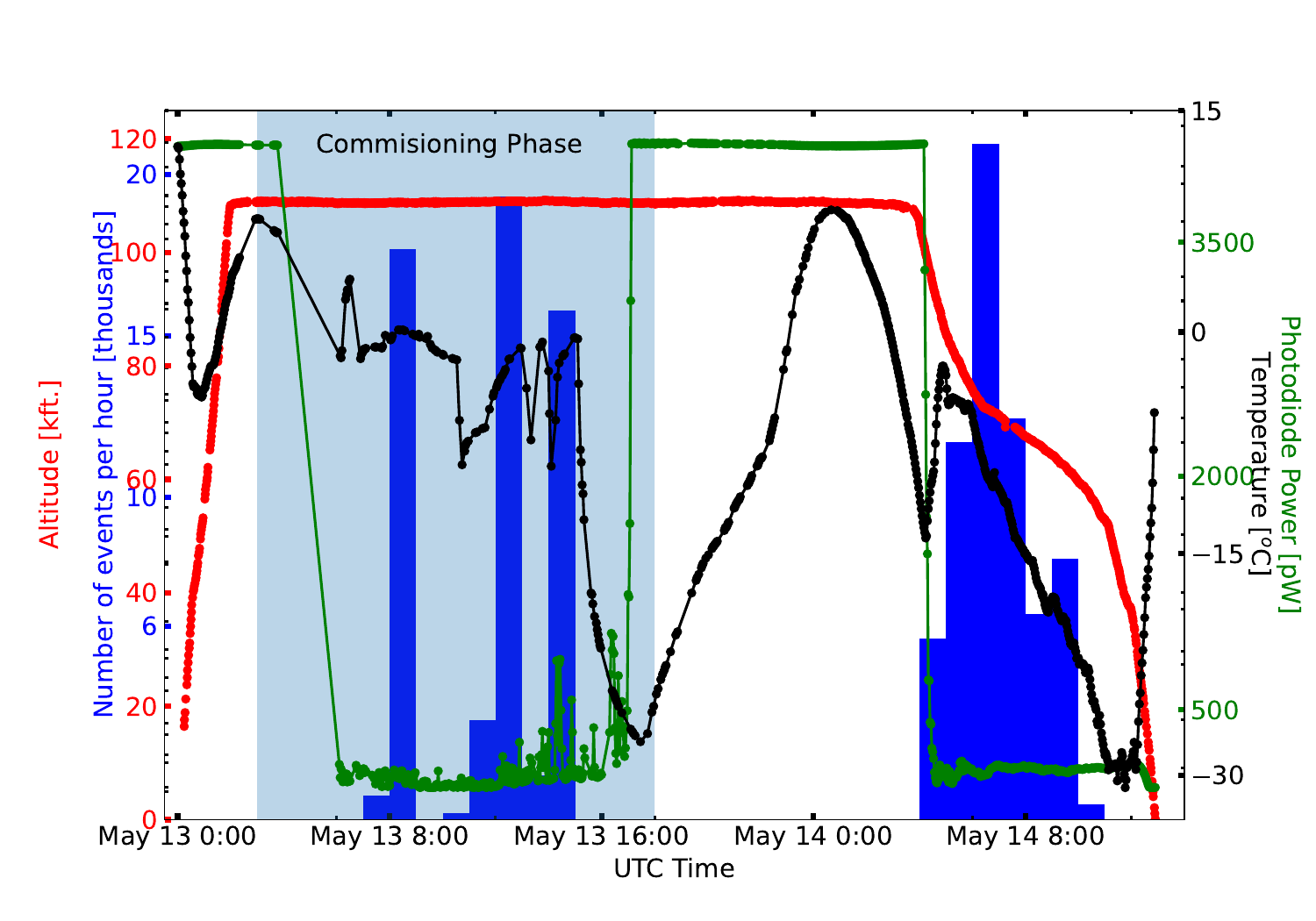}
	\caption{Altitude as a function of time from shortly after launch until flight termination.
	Number of events recorded per hour (blue).
	Temperature Recorded inside of telescope (black),
	light level at focal surface measured by photodiode with a maximum readout of 4,100 picowatts (green). 
	Note: 600 pW is the light level that allows for safe operations of the MAPMTs.  }
	\label{figs:prof}
\end{centering}
\end{figure}

After learning of a leak in the balloon at 02:00 UTC on May 14$^\text{th}$, a plan to address the HVPS issue by bringing the photo-cathode to its nominal voltage more slowly was deployed as rapidly as possible.
By ramping up high voltage in 74 steps taking 148 seconds instead of 8 steps taking 16 seconds, all ECs were able to be turned on successfully. 
After performing a threshold scan (often called an s-curve), the instrument operated for the remainder of the flight. 
With our highest bandwidth telemetry connection, a Starlink maritime unit purchased from SpaceX, fully operational we were able to download the majority of data recorded.
In total 123,781 triggered events were recorded and downloaded. 
While the FT was now operating as expected, the balloon began rapidly decending, until the flight terminated into the southern pacific ocean at 12:54 on May 14$^{\text{th}}$ 2023, 36 hours and 52 minutes after launch.
The time profile of the flight is illustrated in Figure \ref{figs:prof}.

\section{Detector Stability}

A standard diagnostic tool that helps us understand the preformance of our instrument is the "s-curve". 
This is a type of threshold scan, where data is taken while changing the minimum current which must be measured at the anode in order for a count to be recorded. 
At a very low digital analog conversion (DAC) setting, the current required to register one photo-electron is so high that no  photons are counted. 
At a certain value, for our instrument around DAC=820,  the threshold for registering a count is so low  that electronic noise begins to dominate the data. 
Beyond a certain value, the threshold is so low that the dark current never falls below the threshold for a count and as such no counts are recorded. 
S-curves allow for a quick and robust way to gauge the response of the instrument, by both verifying that the thresholds on each pixel are set correctly and measuing the relative efficency of the different parts of the focal surface. 
One s-curve takes roughly three minutes  to record. 

\begin{figure}[h!]\begin{center}
	\includegraphics[width=0.7\paperwidth]{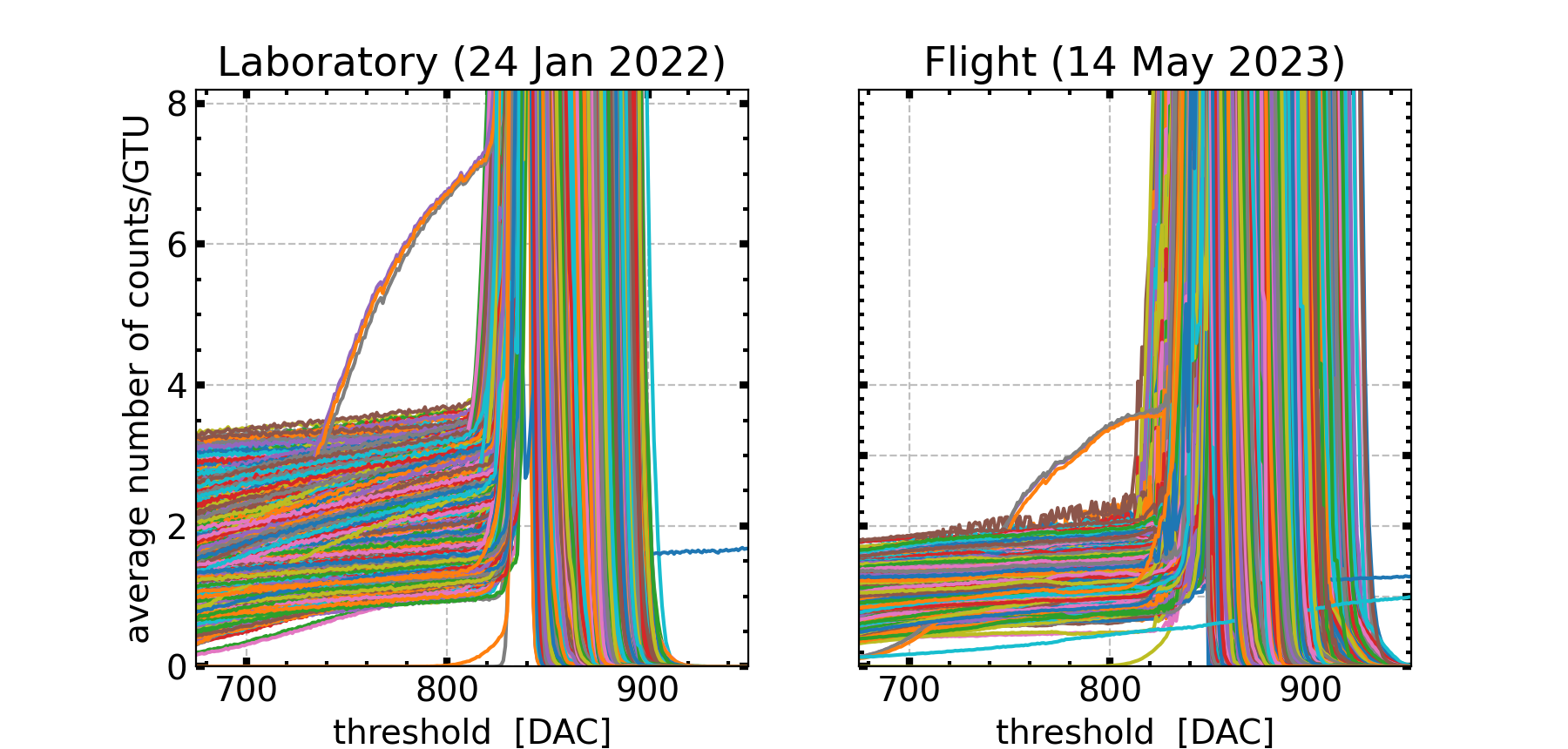}
	\caption{Threshold scan "s-curves" taken during the integration in 2022 (left) and during flight (right). 
	Each line represents a unique pixel, with 6,912 total pixels plotted in each case. 
	Focal surface is illuminated by a controlled light source inside of a dark box (left) and the night sky background(right).
	}
	\label{figs:scurve}
\end{center}\end{figure}

Two sets of s-curve data for all 6,912 channels are shown in Figure \ref{figs:scurve}, one taken on ground laboratory integration in 2022, and one taken during flight. 
As can be seen, several pixels do not behave well in both instances, however the overall response of the detector remains consistent. 

\begin{figure}[h!]\begin{center}
	\includegraphics[width=0.95\textwidth]{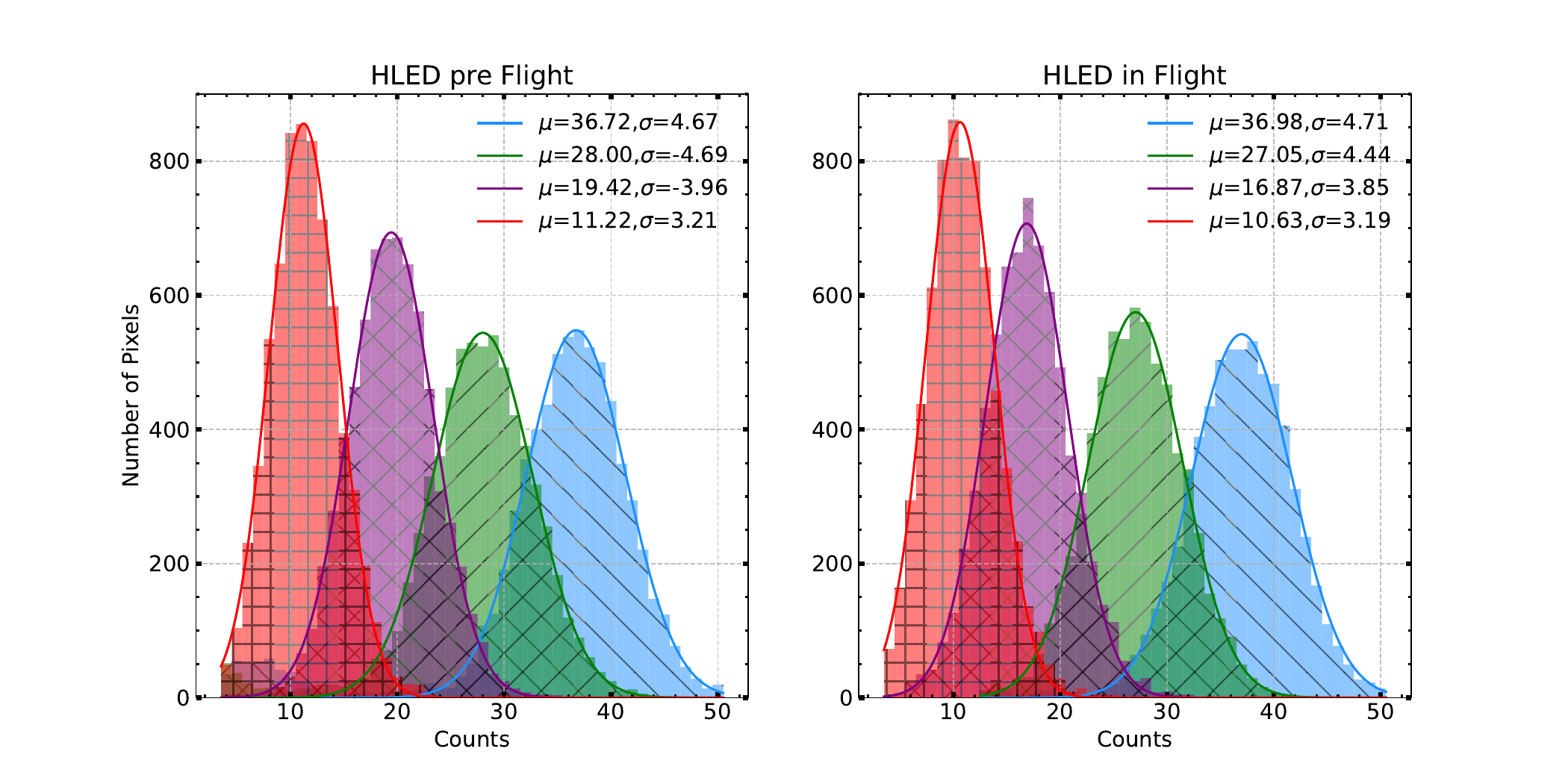}
	\caption{HLED in flight and before. Different flashes, of different intensities, shown in different colors. Solid lines represent three parameter gaussian fits.}
	\label{figs:hled}
\end{center}\end{figure}

Another very useful tool for understanding the performance of the instrument are the HLEDs.
Each of the HLEDs was flashed every 32 seconds, with each flash consisting of a 2.5 $\mu$s pulse and a 5 $\mu$s pulse, seperated by 15 $\mu$s. 
The two HLEDs are offset from one another by 16 seconds, and are programmed to pulse with different intensities. 
Figure \ref{figs:hled} shows the excellent consistency between the data recorded on the ground (left) and in flight (right). 
The frequency of the HLEDs are chosen so that roughly every other file produced, groupings of nine consecutive triggers,  should contain an HLED event. 
By having a total of four different intensities for the HLED, more of the dynamic range of the instrument can be monitored.
Data from the HLEDs are recorded by utilizing the internal triggering system, with no physical link between the HLED systems and the PDMs. 

\section{Exposure Estimation}

The detector's response to expected signals is simulated using the JEM-EUSO OffLine framework \citep{offline}.
Extensive air showers are simulated using the Conex \citep{Bergmann:2006yz} program, an extension of Corsika \citep{corsika}. 
The light generated from these showers is then propagated through the atmosphere, accounting for absorption and scattering. 
A Geant4 implementation of the telescope optics focuses the light onto a parametric focal surface which simulates the response of our electronics with quantum efficiency and double pulse resolution being tuned to laboratory measurements. 
These simulations are validated by data taken in the field with calibrated light sources. 
By isotropically throwing showers over a large area on ground below the simulated detector, the acceptance as a function of energy can be estimated. 
Folding this acceptance in with the spectrum measured by the Pierre Auger Observatory \citep{AugerSpectrum}, allows for an expected event rate to be calculated. 
This event rate can be further constrained by incorporating information from the independent infrared camera \citep{diesing2021ucirc2} which is coaxial to the FT. 
This process is detailed in \citep{Filippatos_2021}.

By accounting for the observational conditions in flight, including cloud coverage, background trigger rate and most significantly the shortened observation period, the expected number of observed EAS is below one.

\section{First Look at Data}

\begin{figure}[h!]\begin{center}
	\includegraphics[width=0.95\textwidth]{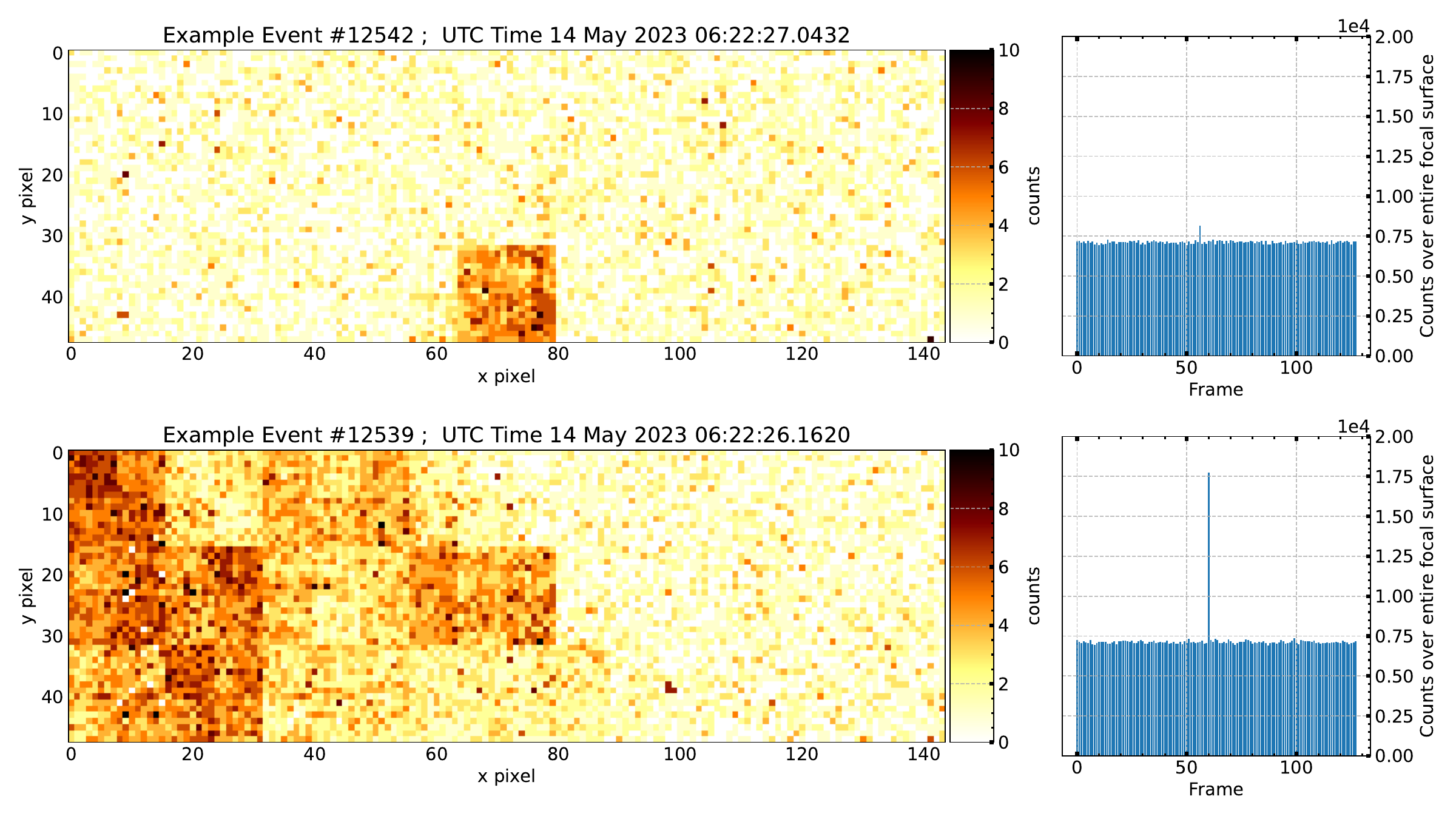}
	\caption{Two example events taken during the second night of the flight. The pattern of increased light follows the layout of the ECs within the focal surface. Time profile of each event shown in the right panels, with the brightest frame shown in the left panels.}
	\label{figs:example}
\end{center}\end{figure}

The majority of events recorded during the short flight were contained in a single 1 $\mu$s frame. 
These events are possibly explained by a charged particle interacting directly with the focal surface causing an excess of hits to be recorded. 
An example of a particularly bright event is shown in Figure \ref{figs:example}, along with an event of typical brightness.
Signal in this example is spread over multiple MAPMTs and two PDMs, however many events are localized to a singular EC.

Track-like EAS signatures have been searched for using a neural network based approach and following a more "traditional" track finding approach using the JEM-EUSO OffLine framework. 
Thus far, no EAS candidates have been found, which is consistent with the expectations given the limited integrated exposure over the short flight. 

\section{Conclusions}

The EUSO-SPB2 FT represents the culmination of over a decade of work within the JEM-EUSO collaboration. 
Functioning at float altitude after an accelerated commissioning phase on a short flight, the MAPMT based fluorescence detector has been shown to be a mature technology.
While the flight did not last nearly as long as was planned hence the integrated exposure of the instrument was greatly limited, the FT was very well positioned to make the first observation of a UHECR air shower from near space. 
The instrument operated near autonomously and in a stable manner through the shortened flight.
Six hours of data were successfully recorded and downloaded with the entire camera functioning nominally.  
Unfortunately, the payload was lost after a bad leak in the balloon forcing a termination after just thirty-six hours aloft.
Further, valuable experience was gained throughout the development and integration of the instrument, during field testing, as well as throughout the flight. 
With the eventual goal of an ultra-wide field-of-view space-based detector, EUSO-SPB2 has provided pioneering insights and can serve as a validation of the technology needed for future missions.

\section*{Acknowledgement}
The authors acknowledge the support by NASA awards 11-APRA-0058, 16-APROBES16-0023, 17-APRA17-0066, NNX17AJ82G, NNX13AH54G, 80NSSC18K0246, 80NSSC18K0473, 80NSSC19K0626, 80NSSC18K0464, 80NSSC22K1488, 80NSSC19K0627 and 80NSSC22K0426, the French space agency CNES, National Science Centre in Poland grant n. 2017/27/B/ST9/02162, and by ASI-INFN agreement n. 2021-8-HH.0 and its amendments. 
This research used resources of the US National Energy Research Scientific Computing Center (NERSC), the DOE Science User Facility operated under Contract No. DE-AC02-05CH11231. 
We acknowledge the NASA BPO and CSBF staffs for their extensive support. 
We also acknowledge the invaluable contributions of the administrative and technical staffs at our home institutions.

\bibliographystyle{JHEP-nt}
\bibliography{references}
\newpage
\newpage
{\Large\bf Full Authors list: The JEM-EUSO Collaboration\\}

\begin{sloppypar}
{\small \noindent
S.~Abe$^{ff}$, 
J.H.~Adams Jr.$^{ld}$, 
D.~Allard$^{cb}$,
P.~Alldredge$^{ld}$,
R.~Aloisio$^{ep}$,
L.~Anchordoqui$^{le}$,
A.~Anzalone$^{ed,eh}$, 
E.~Arnone$^{ek,el}$,
M.~Bagheri$^{lh}$,
B.~Baret$^{cb}$
% ...
}
\end{sloppypar}
\vspace*{.3cm}

{ \footnotesize
\noindent
% ...
$^{cb}$ Universit\'e de Paris, CNRS, AstroParticule et Cosmologie, F-75013 Paris, France\\
% ...
$^{ed}$ Istituto Nazionale di Fisica Nucleare - Sezione di Catania, Italy\\
% ...
$^{eh}$ INAF - Istituto di Astrofisica Spaziale e Fisica Cosmica di Palermo, Italy\\
% ...
$^{ek}$ Istituto Nazionale di Fisica Nucleare - Sezione di Torino, Italy\\
$^{el}$ Dipartimento di Fisica, Universit\`a di Torino, Italy\\
% ...
$^{ep}$ Gran Sasso Science Institute, L'Aquila, Italy\\
% ...
$^{ff}$ Nihon University Chiyoda, Tokyo, Japan\\ 
% ...
$^{ld}$ University of Alabama in Huntsville, Huntsville, AL, USA\\
$^{le}$ Lehman College, City University of New York (CUNY), NY, USA\\
% ...
$^{lh}$ Georgia Institute of Technology, USA\\
% ...
}

\end{document}